\documentstyle[aps,epsf,multicol,psfig]{revtex}
\begin{document}
\draft
\title{Thermal transport through a mesoscopic weak link}
\author{Kelly R. Patton and Michael R. Geller}
\address{Department of Physics and Astronomy, University of Georgia, Athens, 
Georgia 30602-2451}
\date{January 4, 2001}
\maketitle

\begin{abstract}
We calculate the rate of energy flow between two macroscopic bodies, each 
in thermodynamic equilibrium at a different temperature, and joined by 
a weak mechanical link. The macroscopic solids are assumed to be 
electrically insulating, so that thermal energy is carried only by phonons. 
To leading order in the strength of the weak link, modeled here by a 
harmonic spring, the thermal current is determined by a product of the local
vibrational density-of-states of the two bodies at the points of connection.
Our general expression for the thermal current can be regarded as a thermal 
analog of the well-known formula for the electrical current through a 
resistive barrier. It is also related to the thermal Landauer formula in the
weak-tunneling limit. Implications for heat transport experiments on 
dielectric quantum point-contacts are discussed. 
\end{abstract}

\vskip 0.05in
\begin{multicols}{2}

\section{introduction}

Mesoscopic phonon systems are relatively unexplored compared with their 
electronic counterparts. An exception is the recent work on thermal 
conductance quantization in freely suspended one-dimensional dielectric 
wires, where the thermal conductance was found to be $\pi k_{\rm B}^2 T 
/6\hbar$ per transmitted vibrational mode\cite{Schwab etal,Rego and 
Kirczenow}. This behavior parallels the well known electrical conductance 
quantization in units of $e^2 /2 \pi \hbar$ per (spin-resolved) channel 
in one-dimensional mesoscopic conductors\cite{van Wees etal,Wharama 
etal,Beenakker review}. Electrical conductance quantization and many other 
aspects of mesoscopic transport in one-dimensional Fermi liquids, as well as
edge-state transport in integral quantum Hall effect systems, can be 
understood with the Landauer and Landauer-B\"uttiker 
formalisms\cite{Beenakker review,Buttiker review}.

The conventional Landauer formula describes charge transport in mesoscopic 
conductors in the limit where there exists one or more propagating 
channels\cite{Landauer footnote}. Another important transport regime is the 
weak tunneling limit, where the charge conductance is much less than 
$e^2 /2 \pi \hbar$ and, as shown by Schrieffer {\it et al.}\cite{Schrieffer 
etal,Mahan}, is determined by the density-of-states (DOS) obtained from the 
one-particle Green's function.

The thermal analog of the weak tunneling limit has not been addressed 
theoretically and is interesting for several reasons. First, a microscopic
quantum description of thermal conduction through weak links is crucial for 
understanding energy dissipation in nanostructures such as nanoparticles, 
nanotubes, molecular circuits, and nanometer-scale electromechanical 
systems. As we shall demonstrate, the classical theory of thermal 
conduction, based on the heat equation, is entirely inapplicable to these 
systems. Second, thermal conduction through a weak link connected to a 
macroscopic solid turns out to be a sensitive local probe of the surface 
vibrational DOS of that solid, suggesting the possibility of a surface 
microscopy based on a scanning {\it thermal} probe.

In this paper we calculate the rate $I_{\rm th}$ of thermal energy flow 
between two 
macroscopic bodies, each in thermodynamic equilibrium, and joined by a 
weak mechanical link. The weak link may consist of one or more chemical 
bonds, or by a narrow ``neck'' of dielectric material, both of which can be 
accurately modeled by a harmonic spring of stiffness $K$. We obtain a 
general expression for the thermal current that can be regarded as a thermal
analog of the well-known formula, derived by Schrieffer 
{\it et al.}\cite{Schrieffer etal}, for the electrical current through 
a resistive barrier. Our result can also be interpreted as an application 
of the thermal Landauer formula\cite{Rego and Kirczenow,Angelescu etal} in 
the weak tunneling limit, with the energy-dependent phonon transmission 
probability calculated microscopically.

Our work is also related to the classic work of Little\cite{Little} on the
thermal boundary resistance at an interface between two dielectrics, a
solid-solid analog of the Kapitza resistance between solids and superfluid 
Helium caused by phonon scattering at the interface.
A tunneling-Hamiltonian approach similar to ours has been applied to the
Kapitza resistance problem by Sheard and Toombs\cite{Sheard and Toombs}.
In our geometry, however, the thermal resistance comes from scattering at 
the weak link, and the thermal current depends on the elastic properties
of the link and does not vanish if the solids are identical. Heat 
transport in mesoscopic junctions has been studied recently with the 
scattering approach by Cross and Lifshitz\cite{Cross and Lifshitz}. Thermal 
transport through weak links has also been studied in conductors, including 
the two-dimensional electron gas\cite{Molenkamp etal,Kane and Fisher thermal 
FQHE} and one-dimensional Luttinger liquids\cite{Kane and Fisher thermal LL}.

The organization of our paper is as follows: In the next section we describe
in detail our mesoscopic weak-link model, and in Section \ref{local 
vibrational dos section} we define and calculate the local vibrational DOS 
for the macroscopic solids. A general expression for the thermal current is 
derived in Section \ref{thermal current section}. Some experimental 
implications are discussed in Section \ref{nanometer-scale silicon junction 
section}, where we calculate the thermal conductance through a nanometer-scale
junction in Si. Section \ref{discussion section} contains a discussion of the 
differences between electron and phonon tunneling, and also of the role of 
phonon phase coherence in this work.

\section{mesoscopic weak link}
\label{mesoscopic weak link section}

\begin{figure}
\centerline{\psfig{file=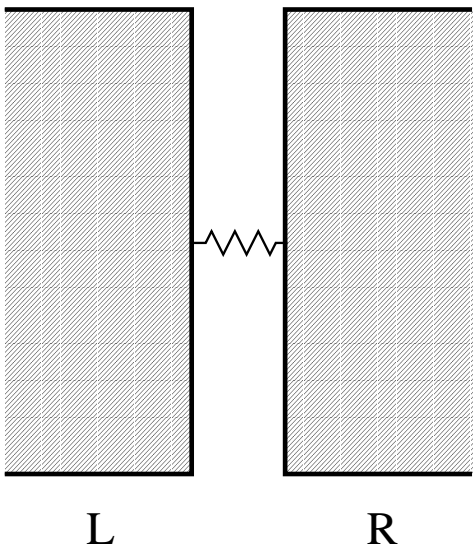,width=2.5in}}
\vspace{0.1in}\setlength{\columnwidth}{3.2in}
\centerline{\caption{Weak link model. Two macroscopic dielectrics, at 
temperatures $T_{\rm L}$ and $T_{\rm R}$, are joined by a harmonic spring of 
stiffness $K$.\label{weak link figure}}}
\end{figure}

The model we consider is as follows: Two macroscopic solids, L and R, are 
held at fixed temperatures $T_{\rm L}$ and $T_{\rm R}$. The two bodies are 
assumed to be electrically insulating, so that thermal energy is carried 
only by phonons. The Hamiltonian of the isolated solids is (we set $\hbar 
= 1$)
\begin{equation}
H_0 = H_{\rm L} + H_{\rm R},
\end{equation}
where
\begin{equation}
H_I \equiv \sum_n \omega_{In} \, a_{In}^\dagger \, a_{In}, \ \ \ \ \ \ \ 
I={\rm L,R}.
\label{unperturbed hamiltonian}
\end{equation}
The $a_{nI}^\dagger$ and $a_{nI}$ are phonon creation and annihilation 
operators for the left and right sides, satisfying
\begin{equation}
[a_{nI}, a_{n'I'}^\dagger] =  \delta_{nn'} \, \delta_{II'}
\end{equation}
and
\begin{equation}
[a_{nI},a_{n'I'}] = [a_{nI}^\dagger, a_{n'I'}^\dagger] = 0.
\end{equation}
The vibrational modes of the isolated bodies are labeled by $n$ and have 
energies $\omega_{In}$. Our analysis is valid for any $\omega_{In}$. The 
mesoscopic weak-link model is illustrated in Fig.~\ref{weak link figure}.

The two macroscopic solids are connected by a weak mechanical link, which we
model by a harmonic spring with stiffness $K$,
\begin{equation}
\delta H = {\textstyle{1 \over 2}} K \big(u_{\rm L}^z - u_{\rm R}^z \big)^2.
\label{interaction definition}
\end{equation}  
Here $u_{I}^z$ is the normal component of the displacement field at the 
surface of body $I$ at the point of connection to the weak link, with the 
surface normal taken to be in the $z$ direction. The surface displacements 
can be expanded in a basis of phonon annihilation and creation operators as
\begin{equation}
u_I^z = \sum_n \big( h_{In} a_{In} + h_{In}^* a_{In}^\dagger \big), \ \ \ \ 
\ \ \ I={\rm L,R}
\label{displacement field expansion}
\end{equation}
where the $h_{In}$ are model-dependent complex coefficients. The $h_{In}$
appropriate for the stress-free planar surface of a semi-infinite isotropic
elastic continuum are given below.

As discussed above, we are interested in systems where the mechanical 
interaction between the two bodies is actually caused by one or more atomic 
bonds, or by a narrow ``neck'' of dielectric material. Our harmonic spring 
model correctly accounts for the longitudinal (normal to the surface) elastic 
forces between the solids, but neglects any transverse or torsional 
interaction. Although transverse and torsional forces can be included by a 
straightforward generalization of our model, they are often much smaller 
than the longitudinal coupling. 

The macroscopic bodies act as thermal reservoirs, and are taken to be ideal 
thermal conductors. In particular, they are assumed to be harmonic [see 
Eq.~(\ref{unperturbed hamiltonian})]. Therefore, the thermal resistance we 
calculate is caused entirely by the scattering of phonons at the junction 
between the reservoirs and the weak link, and by the finite 
transmission probability through the link. 

\section{local vibrational dos}
\label{local vibrational dos section}

In what follows we will need the {\it local} vibrational DOS (or, more
precisely, local spectral density) of the bulk solids, evaluated at the point 
of contact with the weak link. These can be obtained from the retarded 
surface-displacement correlation functions
\begin{equation}
D_I(t) \equiv -i \theta(t) \big\langle [ u_I^z(t) ,  u_I^z(0) ]\big\rangle_0
\label{propagator definition}
\end{equation}
for the isolated macroscopic bodies L and R. Using (\ref{displacement field 
expansion}) leads to
\begin{equation}
D_I(t) = -2 \, \theta(t) \, \sum_n |h_{In}|^2 \sin (\omega_{In}t) .
\end{equation}
The local DOS $N_I(\omega)$ is then defined in terms of the 
Fourier transform of
(\ref{propagator definition}),
\begin{equation}
N_I(\omega) \equiv - {\textstyle{1 \over \pi}} \, {\rm Im} \, D_I(\omega).
\label{local DOS definition}
\end{equation}
Then we have
\begin{equation}
N_I(\omega) = \sum_n |h_{In}|^2 [ \delta(\omega - \omega_{In}) - 
\delta(\omega + \omega_{In}) ].
\end{equation}

In many cases of interest the local spectral density is an algebraic function
of energy at low energies,
\begin{equation}
N_I(\omega) \propto \omega^\alpha,
\label{power-law DOS}
\end{equation}
where $\alpha$ is a constant. For example, $\alpha = 1$ at the planar surface 
of a semi-infinite isotropic elastic continuum (see below).

\section{thermal current}
\label{thermal current section}

We now calculate the heat flow between the two bodies joined by the weak 
link. The complete system is described by the Hamiltonian
\begin{equation}
H = H_0 + \delta H.
\label{full hamiltonian}
\end{equation}
We define a thermal current operator ${\hat I}_{\rm th}$ according to
\begin{equation}
{\hat I}_{\rm th} \equiv \partial_t H_{\rm R} = i[H,H_{\rm R}].
\end{equation} 
The expectation value of ${\hat I}_{\rm th}$ is the energy per unit time 
flowing from the left to the right body.

Writing the interaction (\ref{interaction definition}) as
\begin{equation}
\delta H = {\textstyle{1 \over 2}} K \sum_{nn'} (A_{{\rm L}n} - A_{{\rm R}n})
(A_{{\rm L}n'} - A_{{\rm R}n'}),
\end{equation}
where
\begin{equation}
A_{In} \equiv h_{In} \, a_{In} + h_{In}^* \, a_{In}^\dagger, 
\label{A definition}
\end{equation}
we find that the thermal current operator then takes the form
\begin{equation}
{\hat I}_{\rm th} = {iK \over 2} \sum_{nn'} \omega_{{\rm R}n} \big\lbrace 
A_{{\rm R}n'} - A_{{\rm L}n'} , h_{{\rm R}n} \, a_{{\rm R}n} - 
h_{{\rm R}n}^* \, a_{{\rm R}n}^\dagger \big\rbrace,
\end{equation}
where $\lbrace \cdot \, , \cdot \rbrace$ is an anticommutator.

The equation of motion for the density matrix in the interaction 
representation is
\begin{equation}
\partial_t \rho(t) = i \big[ \rho(t) , \delta H(t) \big],
\label{density matrix equation}
\end{equation}
where
\begin{equation}
O(t) \equiv e^{i H_0 t} O e^{-i H_0 t}.
\end{equation}
From (\ref{density matrix equation}) we find that the nonequilibrium thermal
current to leading order is
\begin{equation}
I_{\rm th}(t) = i \int_0^t dt' \big\langle [ \delta H(t'), 
{\hat I}_{\rm th}(t) ] \big\rangle_0 .
\label{current expectation value}
\end{equation}

Evaluating Eq.~(\ref{current expectation value}) leads to our principal 
result (with factors of $\hbar$ reinstated)
\begin{equation}
I_{\rm th} = {2 \pi K^2 \over \hbar} \int_0^\infty \! d\epsilon \, \epsilon 
\, N_{\rm L}(\epsilon) \, N_{\rm R}(\epsilon) \, \big[ n_{\rm L}(\epsilon)
- n_{\rm R}(\epsilon) \big].
\label{thermal current}
\end{equation}
Here $n_{\rm L}(\epsilon)$ and $n_{\rm R}(\epsilon)$ are Bose distribution 
functions 
\begin{equation}
n(\epsilon) \equiv 1 \big/ \big(e^{\epsilon/k_{\rm B}T} - 1\big)
\label{bose distribution function}
\end{equation}
with temperatures $T_{\rm L}$ and $T_{\rm R}$. The details leading to 
Eq.~(\ref{thermal current}) are given in Appendix \ref{thermal current 
appendix}.

Our result (\ref{thermal current}) shows that the thermal current between
a dielectric held at zero temperature and a second dielectric at temperature
$T$ will be a power-law function of $T$, in striking contrast with 
nonmesoscopic
thermal transport. For example, assuming a spectral density of the form
(\ref{power-law DOS}) leads at low temperature to
\begin{equation}
I_{\rm th} \propto T^{2 \alpha + 2},
\end{equation}
where $T$ is the temperature of the second body.

The linear thermal conductance, defined by 
\begin{equation}
G_{\rm th} \equiv \lim_{T_{\rm L} \rightarrow T_{\rm R}} \ {I_{\rm th} \over
T_{\rm L}-T_{\rm R}},
\end{equation}
is given by
\begin{equation}
G_{\rm th} = {2 \pi K^2 \over \hbar} \int_0^\infty \! d\epsilon \, \epsilon 
\, N_{\rm L}(\epsilon) \, N_{\rm R}(\epsilon) \, { \partial n(\epsilon) 
\over \partial T}.
\label{thermal conductance}
\end{equation}
This expression, along with Eq.~(\ref{power-law DOS}), shows that the linear
thermal conductance between two dielectrics held at a common temperature $T$,
varies at low temperature as a power-law in $T$,
\begin{equation}
G_{\rm th} \propto T^{2\alpha +1},
\label{conductance power-law}
\end{equation}
where $\alpha$ is the exponent characterizing the power-law spectral density
at low energies.

\section{thermal conductance of nanometer-scale silicon junction}
\label{nanometer-scale silicon junction section}

\begin{figure}
\centerline{\psfig{file=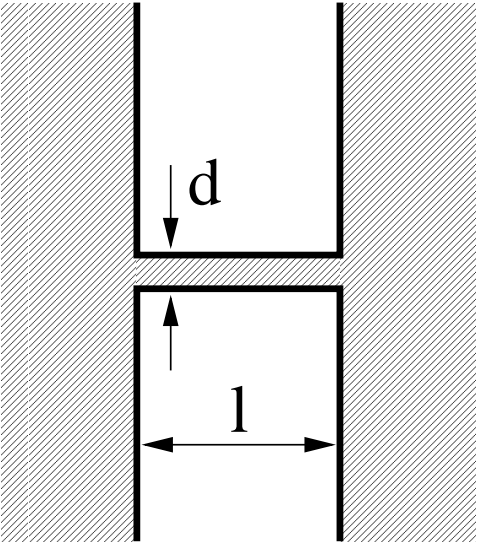,width=2.5in}}
\vspace{0.1in}\setlength{\columnwidth}{3.2in}
\centerline{\caption{Cylindrical silicon junction of length $l$ and diameter
$d$.\label{silicon junction figure}}}
\end{figure}

In this section we give a simple application of our theory to a structure
consisting of a cylindrical neck of Si material connecting two semi-infinite 
Si crystals. To be in the mesoscopic regime we assume the dimensions of
the weak link to be smaller than the phase-coherence length of the 
phonons. The geometry of the system we consider is shown schematically
in Fig.~\ref{silicon junction figure}.

To apply our formula (\ref{thermal conductance}) we need the phonon spectral
density at the surface of Si, and also the effective spring constant of the 
link. The spectral density at energies much less than the Debye energy may be 
obtained from elasticity theory. This approach, which requires a detailed 
consideration of the vibrational modes of a semi-infinite elastic continuum
with a stress-free planar surface, is carried out in Appendix \ref{surface 
DOS appendix}. We show there that the spectral density at the surface of Si is
\begin{equation}
N(\epsilon) = C \epsilon, \ \ \ \ \ C \approx 1.3 \times 10^{8} \
{\rm cm}^2  \, {\rm erg}^{-2}.
\label{Si DOS}
\end{equation}
Then using (\ref{thermal conductance}) we obtain\cite{integral footnote}
\begin{equation}
G_{\rm th} = (8 \pi^5 K^2 C^2 k_{\rm B}^4/15 \hbar) \, T^3.
\label{Si conduuctance formula}
\end{equation}
The longitudinal stiffness of the mechanical link, a cylinder of length
$l$ and diameter $d$, is
\begin{equation}
K = {\pi d^2 \over 4 \, l} \, Y,
\label{effective stiffness}
\end{equation}
where $Y$ is Young's modulus. For Si,\cite{DST book}
\begin{equation}
Y \approx 1.3 \times 10^{12} \, {\rm dyn \ cm^{-2}},
\end{equation}
and assuming link dimensions of $l = 10 \, {\rm nm}$ and $d = 1 \, {\rm nm}$,
we obtain
\begin{equation}
K \approx 1.0 \times 10^{4} \, {\rm erg \ cm^{-2}},
\end{equation}
and a thermal conductance of
\begin{equation}
G_{\rm th} = (9.5 \times 10^{-11} \, {\rm erg \ s^{-1} \ K^{-4}}) \ T^3.
\label{Si conductance}
\end{equation}

It is interesting to compare this result with the ``classical'' thermal 
conductance 
\begin{equation}
G_{\rm th}^{\rm cl} = {\pi d^2 \over 4 \, l} \, \kappa
\label{classical Si conductance}
\end{equation}
of the cylindrical link, as predicted by the heat equation. Here $\kappa$ is
the experimentally measured bulk thermal conductivity, itself a function of
temperature, which for Si can be parameterized as\cite{Fulkerson etal}
\begin{equation}
\kappa =  { 10^7 \, {\rm erg \ s^{-1} \ cm^{-1} \ K^{-1}} \over
0.16 + 1.5 \! \times \! 10^{-3} \, T + 1.6 \! \times \! 10^{-6} \, T^2},
\label{parameterization}
\end{equation}
with temperature in $K$. Eq.~(\ref{parameterization}) is accurate down to
about 100K, below which one can use the low-temperature data of 
Ref.~\cite{Glassbrenner etal}.

\begin{figure}
\centerline{\psfig{file=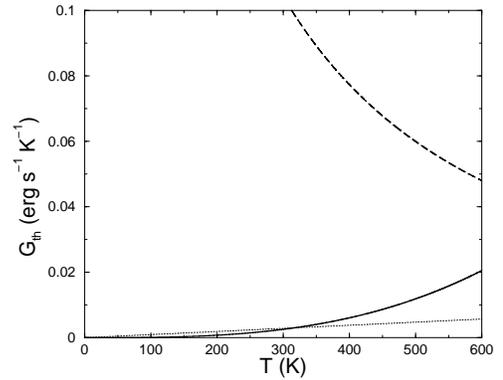,width=2.5in}}
\vspace{0.1in}\setlength{\columnwidth}{3.2in}
\centerline{\caption{Thermal conductance of the mesoscopic Si link shown in
Fig.~\ref{silicon junction figure}, as a function of temperature. The solid
line follows from Eq.~(\ref{Si conductance}), and the dashed line from the
corresponding classical result given in Eq.~(\ref{classical Si conductance}).
The thin dotted line is the universal thermal conductance $\pi k_{\rm B}^2 
T / 6 \hbar$ of a single propagating channel.
\label{conductance figure}}}
\end{figure}

In Fig.~\ref{conductance figure} we plot our result (\ref{Si conductance}) 
along with the classical thermal conductance (\ref{classical Si conductance})
of the same Si link. As is evident, these are dramatically different at low 
temperatures. The large difference occurs because, as discussed in Section
\ref{discussion section}, the origin of the thermal resistance in the two 
formulas (\ref{Si conductance}) and (\ref{classical Si conductance}) are 
different. Although it is tempting to conclude that they begin to agree 
at higher temperature, this would be incorrect, because our theory breaks down
at higher temperature and Eq.~(\ref{Si conductance}) is not valid up to the 
temperature where the curves meet.

There are four reasons why our analysis becomes invalid as the temperature
is increased: The first is that we have assumed the weak link to be of 
mesoscopic dimensions. As the temperature increases, anharmonic interaction 
will eventually make the phonon phase-coherence length smaller than the size 
of the link. While an estimate of the phase-coherence length is beyond the
scope of this work, the experiment of Schwab {\it et al.}\cite{Schwab etal}
suggests that in Si it is at least 1 nm at 1 K. The second is that our 
estimate of the spectral density is only valid for temperatures much less than
the Debye temperature of Si, about 625 K. The third reason is that the 
leading-order perturbation theory we have used breaks down when $G_{\rm th}$ 
approaches the thermal conductance $\pi k_{\rm B}^2 T / 6 \hbar$ corresponding
to one propagating channel, shown as a thin dotted line in 
Fig.~\ref{conductance 
figure}. And the fourth reason is that we have neglected any electronic 
contribution to $G_{\rm th}$, which is correct only when $k_{\rm B} T$ is much
less than the Si band gap. Taking all of these factors into consideration 
suggests that Eq.~(\ref{Si conductance}) is probably not quantitatively 
correct beyond about 10 K.

\section{discussion}
\label{discussion section}

In this paper we have studied the thermal analog of the weak tunneling
limit of charge conduction---which might be regarded as phonon 
``tunneling''---and find many similarities to electron tunneling. 
There are a few important
differences, however.

Electron tunneling, as it is usually defined, involves the passage of an 
electron through a classically forbidden region. In the thermal case, a phonon
of energy $\epsilon$ is never in a region that does support a mode at that 
energy.\cite{tunneling footnote} For example, the harmonic spring employed in 
Eq.~(\ref{interaction definition}) can support a single propagating phonon
channel, but phonons incident on the weak link are mostly reflected back into
the macroscopic dielectrics. Whereas the tunneling rate of an electron (at a 
fixed energy $\epsilon$) through a forbidden region of thickness $l$ varies 
exponentially with $l$ in the weak-tunneling limit, the thickness dependence 
in the phonon tunneling case is different. In the example discussed in Section
\ref{nanometer-scale silicon junction section}, the thermal conductance varies
with length $l$ of the bridge as $l^{-2}$, because the effective spring 
constant (\ref{effective stiffness}) of the bridge becomes softer with 
increasing $l$.

We have demonstrated that the classical theory of thermal conduction, based 
on the heat equation and on the concept of a local thermal conductivity, is 
entirely inapplicable to mesoscopic dielectrics. In a mesoscopic dielectric, 
thermal resistance is caused by elastic scattering of phonons, whereas in an 
infinite, disorder-free crystal it is caused by inelastic scattering due to 
anharmonicity. In the
example of Section \ref{nanometer-scale silicon junction section}, the quantum
result (\ref{Si conduuctance formula}) is determined by the {\it mechanical} 
properties of the bridge material, through the elastic modulus $Y$, whereas 
the classical result (\ref{classical Si conductance}) is determined by the 
bridge material's bulk thermal conductivity $\kappa$.

\acknowledgements

This work was supported by the Research Corporation. It is a pleasure to 
thank Steve Lewis for useful discussions. 

\end{multicols}

\appendix

\section{general formula for the thermal current}
\label{thermal current appendix}

Evaluating (\ref{current expectation value}) we find
\begin{eqnarray}
I_{\rm th}(t) = {K^2\over 2} \sum_{nn'mm'} \int_0^t dt' \ \omega_{{\rm R}n} 
\ {\rm Re} \, &\bigg\langle& \! \bigg\lbrace A_{{\rm L}m}(t') - A_{{\rm R}m}
(t'), \bigg[ A_{{\rm L}m'}(t') - A_{{\rm R}m'}(t') \, , \bigg( A_{{\rm L}n'}
(t) - A_{{\rm R}n'}(t) \bigg) \nonumber \\
&\times&  \bigg( h_{{\rm R}n} \, a_{{\rm R}n}(t) -  h_{{\rm R}n}^* \, 
a_{{\rm R}n}^\dagger(t) \bigg) \bigg] \bigg\rbrace \bigg\rangle_{\!0},
\end{eqnarray}
and, after further simplification,
\begin{eqnarray}
I_{\rm th}(t) &=& {K^2\over 2} \sum_{nn'} \int_0^t dt' \ \omega_{{\rm R}n} 
\ {\rm Re} \ \bigg\langle \! \bigg( \big\lbrace A_{{\rm L}n'}(t) , 
A_{{\rm L}n'}(t') \big\rbrace + \big\lbrace A_{{\rm R}n'}(t) , A_{{\rm R}n'}
(t') \big\rbrace \bigg) \bigg[ h_{{\rm R}n} \, a_{{\rm R}n}(t) - 
h_{{\rm R}n}^* \, a_{{\rm R}n}^\dagger(t) , \, A_{{\rm R}n}(t') \bigg] 
\nonumber \\ &+&  \bigg\lbrace h_{{\rm R}n} \, a_{{\rm R}n}(t) - 
h_{{\rm R}n}^* \, a_{{\rm R}n}^\dagger(t) , \, A_{{\rm R}n}(t')  
\bigg\rbrace \, \bigg( \big[ A_{{\rm L}n'}(t), A_{{\rm L}n'}(t') \big] + 
\big[ A_{{\rm R}n'}(t), A_{{\rm R}n'}(t') \big] \bigg)\bigg\rangle_{\! \! 0},
\end{eqnarray}
where we have used the fact that the commutators are c-numbers. The required
thermal expectation values are
\begin{equation}
\big\langle \big\lbrace A_{In}(t), A_{In}(t') \big\rbrace \big\rangle_0 = 
2 \, |h_{In}|^2 \, \big[1+2 \, n_I(\omega_{In})\big] \cos\omega_{In}(t-t') ,
\end{equation}
\begin{equation}
\big\langle \big[ A_{In}(t), A_{In}(t') \big] \big\rangle_0 = 
-2i \, |h_{In}|^2 \sin\omega_{In}(t-t'), 
\end{equation}
\begin{equation}
\big\langle \big[ h_{{\rm R}n} \, a_{{\rm R}n}(t) - h_{{\rm R}n}^* \, 
a_{{\rm R}n}^\dagger(t) , \, A_{{\rm R}n}(t') \big] \big\rangle_0 = 
2 \, |h_{{\rm R}n}|^2 \cos\omega_{{\rm R}n}(t-t'), 
\end{equation}
and
\begin{equation}
\big\langle \big\lbrace h_{{\rm R}n} \, a_{{\rm R}n}(t) - h_{{\rm R}n}^* \, 
a_{{\rm R}n}^\dagger(t) , \, A_{{\rm R}n}(t') \big\rbrace \big\rangle_0 = 
-2i \, |h_{{\rm R}n}|^2 \, \big[1+2 \, n_{\rm R}(\omega_{{\rm R}n})\big] 
\sin\omega_{{\rm R}n}(t-t').
\end{equation}
These lead to
\begin{eqnarray}
I_{\rm th}(t) = 2 K^2 \sum_{nn'} \ &\omega_{{\rm R}n}& \ |h_{{\rm R}n}|^2 
\ \int_0^t dt' \big( \, |h_{{\rm L}n'}|^2 \big[1+2 \, n_{\rm L}(\omega_{{\rm
L}n'})\big] 
\cos\omega_{{\rm L}n'}(t-t') \, \cos\omega_{{\rm R}n}(t-t') \nonumber \\
&+& |h_{{\rm R}n'}|^2 \big[1+2 \, n_{\rm R}(\omega_{{\rm R}n'})\big] 
\cos\omega_{{\rm R}n}(t-t') \, \cos\omega_{{\rm R}n'}(t-t') \nonumber \\
&-& |h_{{\rm L}n'}|^2 \big[1+2 \, n_{\rm R}(\omega_{{\rm R}n})\big] 
\sin\omega_{{\rm L}n'}(t-t') \, \sin\omega_{{\rm R}n}(t-t') \nonumber \\
&-& |h_{{\rm R}n'}|^2 \big[1+2 \, n_{\rm R}(\omega_{{\rm R}n})\big] 
\sin\omega_{{\rm R}n}(t-t') \, \sin\omega_{{\rm R}n'}(t-t') \big) .
\end{eqnarray}
Here $n_{\rm L}(\epsilon)$ and $n_{\rm R}(\epsilon)$ are Bose 
distribution functions [see Eq.~(\ref{bose distribution function})] 
with temperatures $T_{\rm L}$ and $T_{\rm R}$. Next we make a change of 
variables $t' \rightarrow t-t'$, take the $t \rightarrow \infty$ limit, and 
include a convergence factor to regularize the long-time behavior of the 
resulting integrals. Finally, using the identities
\begin{equation}
\int_0^\infty \! dt \, \cos\omega t \, \cos\omega't \ e^{-\zeta t} = {\pi 
\over 2} \big[ \delta(\omega - \omega') + \delta(\omega + \omega') \big] 
\end{equation}
and
\begin{equation}
\int_0^\infty \! dt \, \sin\omega t \, \sin\omega't \ e^{-\zeta t} = {\pi 
\over 2} \big[ \delta(\omega - \omega') - \delta(\omega + \omega') \big],
\end{equation}
\begin{multicols}{2}
\noindent where $\zeta$ is a positive infinitesimal, and reinstating factors
of $\hbar$, leads to Eq.~(\ref{thermal current}). 

In Eqs.~(\ref{thermal current}) and (\ref{thermal conductance}) we have 
introduced an {\it energy}-dependent DOS,
\begin{equation}
N_I(\epsilon) \equiv \sum_n |h_{In}|^2 [ \delta(\epsilon - \hbar 
\omega_{In}) - \delta(\epsilon + \hbar \omega_{In}) ]
\label{energy dependent DOS}
\end{equation}
which has dimensions of (length)$^2$/energy. In a homogeneous elastic 
continuum of mass density $\rho$ and volume $V$, $N(\epsilon)$ is equal to 
$\hbar^2 / 2 \rho \epsilon$ times the thermodynamic DOS per volume,
$V^{-1} \sum_n \delta(\epsilon - \epsilon_n).$

\section{surface DOS of silicon}
\label{surface DOS appendix}

In this appendix we calculate the local phonon DOS at the stress-free planar 
surface of a semi-infinite isotropic elastic continuum, following closely the 
work of Ezawa \cite{Ezawa}, and use this to estimate the DOS at the surface of
Si. The substrate is assumed to occupy the space $z \ge 0$. The vibrational 
modes are labeled by $n = (m, {\bf K}, c),$
where $m$ is a branch index taking the values SH, $\pm$, 0, and R, ${\bf K}$
is a two-dimensional wave vector in the $xy$ plane, and $c \equiv \omega /
|{\bf K}|$ is a parameter (continuous for all branches except $m= {\rm R}$) 
with dimensions of velocity. In contrast to Ref.~\onlinecite{Ezawa} we shall 
use periodic boundary conditions in the $x$ and $y$ directions, over a square 
of area ${\cal A}$.

In our analysis we will approximate Si as an isotropic elastic continuum
with longitudinal and transverse sound velocities
\begin{eqnarray}
v_{\rm l} &=& 8.5 \times 10^5 \, {\rm cm \ s}^{-1}, \nonumber \\
v_{\rm t} &=& 5.9 \times 10^5 \, {\rm cm \ s}^{-1},
\label{Si velocities}
\end{eqnarray}
and mass density
\begin{eqnarray}
\rho = 2.3  \, {\rm g \ cm}^{-3}.
\label{Si density}
\end{eqnarray}

It will be convenient to treat the Rayleigh branch ($m = {\rm R}$) 
separately, and then consider the branches with continuous $c$. In the 
Rayleigh case the displacement field is expanded as\cite{R footnote}
\begin{equation}
{\bf u} = \sum_{\bf K} \sqrt{\hbar \over 2 \rho c_{\rm R} |{\bf K}|}
\, \big[ a_{{\rm R}{\bf K}} \, {\bf f}_{{\rm R}{\bf K}} +
a_{{\rm R}{\bf K}}^\dagger \, {\bf f}_{{\rm R}{\bf K}}^*\big],
\end{equation}
where the vibrational eigenfunctions ${\bf f}_{{\rm R}{\bf K}}({\bf r})$ have
dimensions of $L^{-{3 \over 2}}$ and satisfy
\begin{equation}
\int d^3r \ {\bf f}_{{\rm R}{\bf K}}^* \cdot {\bf f}_{{\rm R}{\bf K}'} = 
\delta_{\bf K K'}.
\end{equation}
Here $c_{\rm R}= \xi \, v_{\rm t}$, where $\xi$ is the root between 0 and 1 of
\begin{equation}
\xi^6 - 8 \xi^4 + 8(3-2\nu^2) \xi^2 - 16(1-\nu^2) = 0,
\end{equation}
and where
\begin{equation}
\nu \equiv v_{\rm t}/ v_{\rm l}
\end{equation}
is the ratio of transverse and longitudinal bulk sound velocities. For Si,
$\nu = 0.69$ and $\xi = 0.88$; hence
\begin{eqnarray}
c_{\rm R} &=& 5.2 \times 10^5 \, {\rm cm \ s}^{-1}.
\label{Si surface velocity}
\end{eqnarray}

The $z$ component of the vibrational eigenfunction at the point ${\bf r}=0$ on
the surface is
\begin{equation}
f^z_{\rm R}(0) = \sqrt{ 2 \gamma^3 \eta^2 |{\bf K}|
\over (\gamma-\eta)(\gamma - \eta + 2 \gamma \eta^2) {\cal A} }
\bigg[ 1 - \bigg({2\over 1 + \eta^2} \bigg) \bigg],
\end{equation}
where
\begin{equation}
\gamma \equiv \sqrt{1-(c_{\rm R}/v_{\rm l})^2} \ \ \ {\rm and} \ \ \
\eta \equiv  \sqrt{1-(c_{\rm R}/v_{\rm t})^2}.
\end{equation}
We find that the Rayleigh branch contributes to the local DOS (\ref{local 
DOS definition}) an amount (for positive $\omega$)
\begin{equation}
{\rm R \ branch}: \ \ \ \ \ \ N(\omega) = {g_1 \hbar \omega \over 4 \pi 
\rho c_{\rm R}^3}, \ \ \ \ g_1 \approx 0.42.
\label{R branch DOS}
\end{equation} 
Note that $g_1$ generally depends on $\nu$, the value quoted in (\ref{R branch
DOS}) corresponding to Si.

Next we consider the branches with continuous $c$. Here 
\begin{equation}
{\bf u} = \sum_{\bf K} \int_\Gamma dc \ \sqrt{\hbar \over 2 \rho c  
|{\bf K}|} \, \big[ a_{m{\bf K}c} \, {\bf f}_{m{\bf K}c} +
a_{m{\bf K}c}^\dagger \, {\bf f}_{m{\bf K}c}^* \big],
\label{continuous c displacement field definiton}
\end{equation}
where the vibrational eigenfunctions have dimensions of $L^{-{3 \over 2}}
c^{-{1 \over 2}}$ and satisfy
\begin{equation}
\int d^3r \ {\bf f}_{m{\bf K}c}^* \cdot {\bf f}_{m'{\bf K}'c'} = \delta_{mm'}
\, \delta_{\bf K K'} \, \delta(c-c').
\end{equation}
The range $\Gamma$ of the $c$ integration in (\ref{continuous c displacement 
field definiton}) is $[v_{\rm t},\infty]$ for $m={\rm SH}$, $[v_{\rm l},
\infty]$ for $m=\pm$, and $[v_{\rm t},v_{\rm l}]$ for the $m=0$ branch.
The contribution to the local DOS (for $\omega \ge 0$) from these branches
is given by
\begin{equation}
N(\omega) = {\hbar \over 2 \rho \omega} \sum_{\bf K} \int_\Gamma dc \
|f_{m{\bf K}c}^z(0)|^2 \, \delta(\omega - c |{\bf K}|).
\label{continuous DOS expression}
\end{equation}

The SH modes are polarized in the $xy$ plane and therefore do not contribute
to (\ref{local DOS definition}). The $\pm$ modes have surface amplitude
\begin{eqnarray}
f^z_{\pm}(0) &=& \sqrt{ |{\bf K}| \over 4 \pi c {\cal A}}
\, \bigg[ \pm \sqrt{\alpha} \, (1+A \pm i B) \nonumber \\
&+& {i \over \sqrt{\beta}}(1 - A \mp i B) \bigg],
\end{eqnarray}
where
\begin{equation}
A\equiv { (\beta^2-1)^2 - 4 \alpha \beta \over (\beta^2-1)^2 + 4 \alpha
\beta},
\end{equation}
\begin{equation}
B\equiv { 4 \sqrt{\alpha \beta} (\beta^2-1) \over (\beta^2-1)^2 + 4 \alpha
\beta},
\end{equation}
\begin{equation}
\alpha\equiv \sqrt{(c/v_{\rm l})^2 -1}, \ \ \ {\rm and} \ \ \ 
\beta \equiv \sqrt{(c/v_{\rm t})^2 -1},
\end{equation}
are all real functions of $c$. The $m=\pm$ branches together contribute an 
amount
\begin{equation}
\pm \ {\rm branches}: \ \ \ \ \ \ N(\omega) = {g_2 \hbar \omega \over 4 \pi^2 
\rho v_{\rm l}^3}, \ \ \ \ g_2 \approx 1.0.
\label{pm branch DOS}
\end{equation}
The value for $g_2$, obtained by doing the integration over $c$ in
Eq.~(\ref{continuous DOS expression}) numerically, is valid only for the
value of $\nu$ corresponding to Si.

The $m=0$ branch has amplitude
\begin{equation}
f^z_0(0) = \sqrt{|{\bf K}| \over 2 \pi c \beta {\cal A}}
\, \big[ - \gamma D + i(1-E) \big],
\end{equation}
where
\begin{equation}
D \equiv {4 \beta (\beta^2-1)^3 - 16 i \gamma \beta^2 (\beta^2-1) \over 
(\beta^2-1)^4 + 16 \gamma^2 \beta^2}
\end{equation}
and
\begin{equation}
E \equiv  {(\beta^2-1)^4 - 16 \gamma^2 \beta^2 - 8 i \gamma \beta
 (\beta^2-1)^2 \over 
(\beta^2-1)^4 + 16 \gamma^2 \beta^2}
\end{equation}
are complex-valued functions of $c$. This leads to a contribution
\begin{equation}
{\rm 0 \ branch}: \ \ \ \ \ \ N(\omega) = {g_3 \hbar \omega \over 8 \pi^2 
\rho v_{\rm t}^3}, \ \ \ \ g_3 \approx 0.59.
\label{0 branch DOS}
\end{equation} 
As before, $g_3$ is obtained numerically and assumes a value of $\nu$ valid
for Si. Combining the three contributions (\ref{R branch DOS}), (\ref{pm 
branch DOS}), and (\ref{0 branch DOS}), yields\cite{bulk footnote}
\begin{equation}
N(\omega) = {\hbar \omega \over 4 \pi^2 \rho} \bigg[ {g_1 \pi \over 
c_{\rm R}^3} + {g_2 \over v_{\rm l}^3} + {g_3 \over 2 v_{\rm t}^3} \bigg].
\label{surface spectral density}
\end{equation}
Using Eq.~(\ref{surface spectral density}) we obtain the estimate (\ref{Si 
DOS}).

\end{multicols}

\begin{references}

\bibitem{Schwab etal} K. Schwab, E. A. Henriksen, J. M. Worlock, and M. L. 
Roukes, Nature {\bf 404}, 974 (2000).

\bibitem{Rego and Kirczenow} L. G. C. Rego and G. Kirczenow, Phys. Rev. 
Lett. {\bf 81}, 232 (1998).

\bibitem{van Wees etal} B. J. van Wees, H. van Houten, C. W. J. Beenakker, 
J. G. Williamson, L. P. Kouwenhoven, D. van der Marcel, and C. T. Foxen, 
Phys. Rev. Lett. {\bf 60}, 848 (1988).

\bibitem{Wharama etal} D. A. Wharama {\it et al.}, J. Phys. C {\bf 21}, 
L209 (1988).

\bibitem{Beenakker review} For a review see C. W. J. Beenakker and H. van 
Houten in {\it Solid State Physics: Advances in Research and Applications}, 
edited by H. Ehrenreich and D. Turnbull (Academic Press, San Diego, 1991), 
Vol. 44.

\bibitem{Buttiker review} M. B\"uttiker, in {\it Semiconductors and 
Semimetals}, edited by M. Reed (Academic Press, San Diego, 1992), Vol. 35.

\bibitem{Landauer footnote} The Landauer formula can be applied to 
the case of weak tunneling provided one 
includes an energy-dependent transmission probability for each channel. 
However, these transmission probabilities would have to be calculated 
microscopically, which amounts to using our approach.

\bibitem{Schrieffer etal} J. R. Schrieffer, D. J. Scalapino, and J. W. 
Wilkins, Phys. Rev. Lett. {\bf 10}, 336 (1963).

\bibitem{Mahan} See also G. D. Mahan, {\it Many-Particle Physics}, 3rd ed. 
(Plenum Publishers, New York, 2000).

\bibitem{Angelescu etal} D. E. Angelescu, M. C. Cross, and M. L. Roukes, 
Superlattices Microstruct. {\bf 23}, 673 (1998).

\bibitem{Little} W. A. Little, Can. J. Phys. {\bf 37}, 334 (1959).

\bibitem{Sheard and Toombs} F. W. Sheard and G. A. Toombs, J. Phys. C {\bf
5}, L166 (1972).

\bibitem{Cross and Lifshitz} M. C. Cross and R. Lifshitz, cond-mat/0011501.

\bibitem{Molenkamp etal} L. W. Molenkamp, Th. Gravier, H. van Houten, O. J.
A. Buijk, M. A. A. Mabesoone, and C. T. Foxen, Phys. Rev. Lett. {\bf 68},
3765 (1992).

\bibitem{Kane and Fisher thermal FQHE} C. L. Kane and M. P. A. Fisher, Phys.
Rev. B {\bf 55}, 15832 (1997).

\bibitem{Kane and Fisher thermal LL} C. L. Kane and M. P. A. Fisher, Phys. 
Rev. Lett. {\bf 76}, 3192 (1996).

\bibitem{integral footnote} Note that $\int_0^\infty dx \, x^4 e^x (e^x-1)^{-2}
=4 \pi^4/15.$

\bibitem{DST book} {\it Semiconductors: Group IV Elements and III-V Compounds},
edited by O. Madelung (Springer-Verlag, Berlin, 1991).

\bibitem{Fulkerson etal} W. Fulkerson, J. P. Moore, R. K. Williams, R. S.
Graves, and D. L. McElroy, Phys. Rev. {\bf 167}, 765 (1968).

\bibitem{Glassbrenner etal} C. J. Glassbrenner and G. A. Slack, Phys. Rev. 
{\bf 134}, A1058 (1964).

\bibitem{tunneling footnote} This is correct only at the low energies of
interest in the present work. At higher energies it is of course possible to 
have forbidden regions, for example, in a vibrational analog of a photonic 
band-gap structure, or simply at energies higher than the acoustic and optical
phonon bands of an ordinary crystal.

\bibitem{Ezawa} H. Ezawa, Ann. Phys. {\bf 67}, 438 (1971).

\bibitem{R footnote} Here `R' denotes the Rayleigh branch.

\bibitem{bulk footnote} It is useful to compare this result with the 
corresponding {\it bulk} spectral density of an isotropic elastic continuum, 
calculated with periodic boundary conditions,
$$ N_{\rm bulk} (\omega) = {\hbar \omega \over 4 \pi^2 \rho} \bigg( {1 \over 
v_{\rm l}^3} + {2 \over v_{\rm t}^3} \bigg).$$

\end{references}
\end{document}